\documentclass[11pt,twoside]{article}
\usepackage{macro-fsut-eng}
\usepackage{psfig}
\usepackage{graphicx}
\usepackage{cite}
\usepackage[T1]{fontenc} 

\usepackage{latexsym}
\usepackage{verbatim}

\begin{document}

\vskip 1.0cm
\markboth{B.K.~Gibson et~al.}{Gas and Metals in Simulated Disk Galaxies}
\pagestyle{myheadings}

\vspace*{0.5cm}
\title{Gas and Metal Distributions within Simulated Disk Galaxies}

\author{B.K.~Gibson$^{1,2,3}$, S.~Courty$^{4}$, 
D.~Cunnama$^{5}$ and M.~Moll\'a$^{6}$}
\affil{$^1$Dept of Phys \& Astro, Saint Mary's Univ, Halifax, B3H~3C3, Canada\\
$^2$Monash Centre for Astrophysics, Clayton, 3800, Australia\\
$^3$Jeremiah Horrocks Institute, UCLan, Preston, PR1~2HE, UK\\
$^4$Universit\'e Lyon~1, F-69230, France\\
$^5$Physics Dept, Univ of the Western Cape, Cape Town, South Africa\\
$^6$Dept de Investigaci\'on B\'asica, CIEMAT, Madrid, E28040, Spain}

\begin{abstract}
We highlight two research strands related to our ongoing
chemodynamical Galactic Archaeology efforts: (i) the
spatio-temporal infall rate of gas onto the disk, drawing analogies
with the infall behaviour imposed by classical galactic chemical
evolution models of inside-out disk growth; (ii) the radial age 
gradient predicted by spectrophometric models of disk galaxies. 
In relation to (i), at low-redshift, we find that half of the infall
onto the disk is gas associated with the corona, while half can
be associated with cooler gas streams; we also find that gas
enters the disk preferentially orthogonal to the system, rather
than in-plane.  In relation to (ii), we 
recover age gradient troughs/inflections consistent
with those observed in nature, without recourse to radial migrations.
\end{abstract}

\section{Introduction}
\label{intro}

The infall of gas onto galaxies is a fundamental constituent of 
any cosmologically-motivated models of galaxy evolution, whether
they be classic galactic chemical evolution models (e.g. 
Lineweaver et~al. 2004; Renda et~al. 2004) or hydrodynamical
simulations (e.g. Kawata \& Gibson 2003; Brook et~al. 2004).  
The shape of the metallicity distribution function
of a given ensemble of stars can be a powerful tool to constrain the
(otherwise, little known) 
interplay between infalling and outflowing material (e.g. 
Fenner \& Gibson 2003; Pilkington et~al. 2012b).
In the local Universe, we often associate (rightly or wrongly) this 
infalling fuel for future star formation with the high-velocity clouds
which permeate our halo (e.g. Gibson et~al. 2001; Pisano et~al. 2004).

\begin{figure}
\begin{center}
\hspace{0.50cm}
\psfig{figure=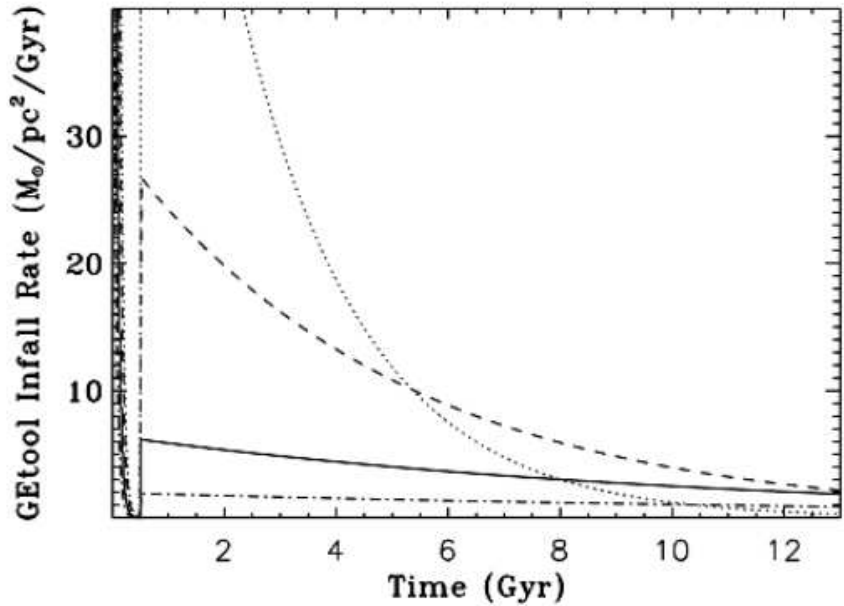,width=12.cm}
\caption{Gas infall rate onto the disk as a function of time ($x$-axis) 
and radius (ranging from inner to outer disk, going from the upper
to the lower curves). The infall timescale here (from the models
of Renda et~al. 2004) grows linearly with galactocentric radius.}
\label{fig1}
\end{center}
\end{figure}

Classic chemical evolution models, constrained by both the metallicity
distribution function and gradients (abundance and surface density)
in the disk, suggest ``inside-out'' growth of the disk is required
(e.g. Chiappini et~al. 2001; Fenner \& Gibson 2003; Moll\'a \& D\'iaz
2005; Pilkington et~al. 2012a,b).  By ``inside-out'', we mean 
a scenario in which the timescale for gas infall onto the disk 
increases as a function of galactocentric radius; whether this
timescale is linear (e.g. Chiappini et~al. 2001) or a higher-order
parametrisation (e.g. Moll\'a \& D\'iaz 2005) is less important than the
fact that (a) the overall infall rate is (roughly) exponential, and
(b) the timescale increases with radius.  Fig~\ref{fig1} shows
one such parametrisation from the chemical evolution model of 
Renda et~al. (2004).

As alluded to above, energetic outflows (driven by a combination of
thermal/kinetic energy from supernovae and massive star winds/radiation)
also play a critical role in regulating star formation, infall of both 
fresh and re-cycled disk material, and setting the chemistry of the
resulting system.  The heating and chemical profiles of the halo are
an ideal place to examine the veracity of one's feedback scheme through 
a comparison of the radial profiles of various neutral and ionised 
species (e.g. \textsc{Hi}, \textsc{Ovi}, \textsc{Mgii}) with those
observed in nature.  We have made significant strides in this area
(Stinson et~al. 2012), and future works in this series will examine
these radial distributions using a statistical sample of cosmolgical
disk simulations in a variety of environments, from field to groups.

The spatio-temporal infall
pattern of gas onto the disk and the predicted 
age gradients within spectrophotometric
disk models are touched upon in the following sub-sections.  These are 
each, very much, works in progress, rather than finished products, so we 
ask the reader to bear that in mind!

\section{How Does Gas Get Into Galaxies?}
\label{infall}

Making use of \textsc{Ramses-CH} (Few et~al. 2012), a
new self-consistent implementation of chemical 
evolution within the \textsc{Ramses} cosmological adaptive
mesh refinement code (Teyssier 2002), we re-simulate the
disk described by S\'anchez-Bl\'azquez et~al. (2009) and analyse the
temporal and spatial infall rates of hot/coronal and cooler/stream gas
onto the disk.  Our task is a (seemingly) straightforward one: confirm/refute
the aforementioned fundamental tenet of chemical evolution, that
the gas infall onto simulated disks (in a cosmological context)
proceeds in an inside-out fashion.

In the upper-left panel of Fig~2, we show in black (magenta)
the inflowing (outflowing)
gas flux through parallel (0.5~kpc thick) slabs situated $\pm$5~kpc
from the mid-plane (extending to a galactocentric radius of 
25~kpc),\footnote{The choice of
$\pm$5~kpc heights is a compromise between being as close to the
disk as possible, without being `swamped' by the galactic
fountain/re-circulation signal (Gibson et~al. 2009; \S7).}
over the range of time for which the disk
could be `readily' identified (see S\'anchez-Bl\'azquez et~al. 2009 for
details pertaining to the `disk identification').  
In the upper-right panel of Fig~2, we decompose the infalling gas
(black curve from the left panel, repeated again here, also in
black) into polytropic/hot gas (what we label as `corona') in
red, and non-polytropic/cooler gas (what we label as `streams') in blue.
Within this simulation, \it (i) the infall from the corona is roughly constant
in time, and (ii) at low-redshift, 
each component accounts for half of the current gas infall. \rm  
In the lower-left panel of Fig~2, we show the infalling (outflowing) gas
flux, again in black (magenta), through a cylinder of radius 25~kpc and
height $\pm$5~kpc; i.e., the sum of the fluxes shown here, plus those
shown in the upper-left panel of Fig~2, correspond to the real/total
accretion rate.  For this simulation, \it the rate of gas infall/inflow
entering the disk through the cylinder (i.e., `in-plane' infall/inflow)
is fairly negligible. \rm
Finally, in the lower-right panel of Fig~2, we show the gas inflow rate through
three small `holes' situated $\pm$5~kpc from the mid-plane, at different
radii.  While noisy, due to the small sampling employed here, we can
see the emergence of the fundamental tenet of inside-out disk growth:
specifically, the lack of gas accretion at small galacto-centric radii
at low-redshift (i.e., a trend for flatter infall profiles at larger radii).

\begin{figure}
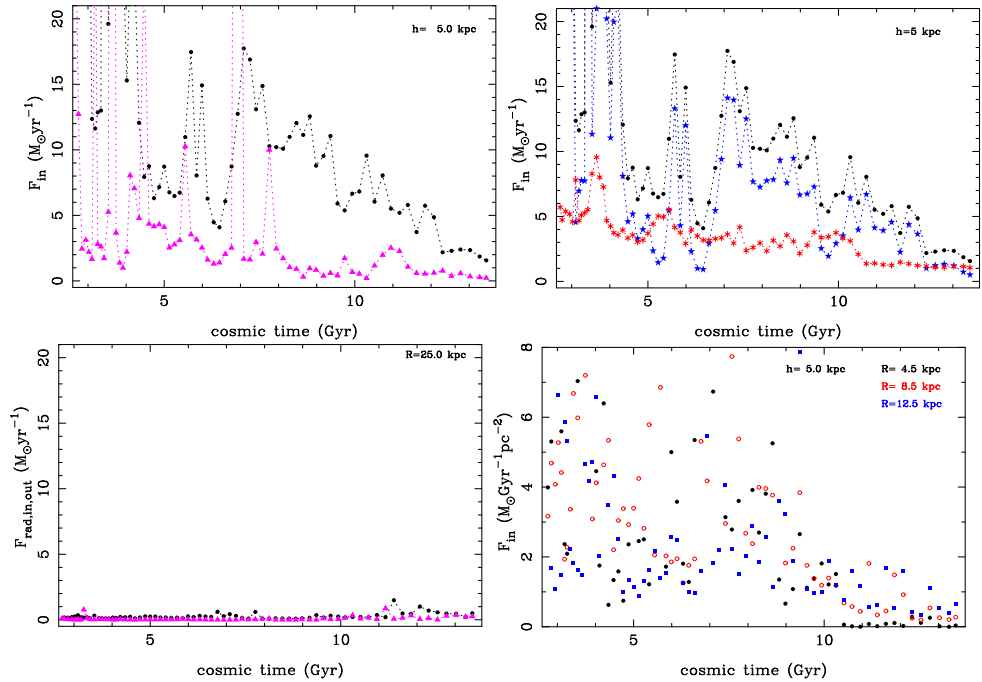

\begin{center}
\hspace{0.25cm}
\psfig{figure=Fig2a.ps,angle=-90,width=6.3cm}
\psfig{figure=Fig2b.ps,angle=-90,width=6.3cm}
\psfig{figure=Fig2c.ps,angle=-90,width=6.3cm}
\psfig{figure=Fig2d.ps,angle=-90,width=6.3cm}
\caption{Upper-left: inflowing (outflowing) gas through slabs
$\pm$5~kpc from the mid-plane in black (magenta); Upper-right: inflowing
hot/coronal (cooler/stream) gas through the same $\pm$5~kpc slabs in
red (blue); Lower-left: inflowing (outflowing) gas through a
$\pm$5~kpc high cylinder of radius 25~kpc in black (magenta);
Lower-right: inflowing gas through small `holes' $\pm$5~kpc
from the mid-plane, at three different radii. See text for details.
}
\label{fig3}
\end{center}
\end{figure}

\section{Age Gradients}
\label{age}

The existence of U-shaped radial age profiles (inferred via radial
colour profiles, in consort with stellar population modeling)
in disk galaxies (particularly those with so-called
Type~II surface brightness
profiles - i.e., those showing a `break' in the surface
brightness) is now well-established (e.g. Bakos et~al. 2008; 
S\'anchez-Bl\'azquez et~al. 2011; Roediger et~al. 2012).  Such troughs 
in age, near the break radius, were found in the exquisite models of
Roskar et~al. (2008), where the `up-bend' in the age profile in the outer
disk was produced by stars that had migrated from the inner parts of the
disk; in our cosmological simulation (S\'anchez-Bl\'azquez et~al. 2009), 
a similar trough/inflection in the age profile was found at the break radius,
where the presence of a warp in the gas disk resulted in a decrease in 
the volume density and, hence, a `break' in the star formation density.
The trough persists, regardless of the presence (or lack thereof) of
radial migration (although migration \it clearly \rm takes place
and \it is \rm critical!).

Whether U-shaped age profiles are a natural byproduct within classical
galactic chemical evolution models is less certain; to that end, we
examined the spectrophotometric predictions associated with the 
same fiducial Milky Way models (Moll\'a \& D\'iaz 2005; N=28) that
were employed in our reecnt work on the temporal evolution of 
metallicity gradients in L$^\star$ galaxies (Pilkington et~al. 2012a).
In Fig~3, we show predicted present-day mass-weighted age gradients
for a Milky Way analog, employing a range of star formation efficiencies
(from high efficiency to low efficiency, in going from the top to the
bottom curves at small galactocentric radii).  We find that within
these classical models, which by construct neglect radial migration, 
U-shaped age profiles are a natural outcome of the infall/star formation
prescriptions.  It is interesting to note that the position and
depth of the trough is sensitive to the adopted star formation efficiency;
high efficiencies drive the trough to be (i) positioned at
larger galactocentric radii, and (ii) shallower (and vice versa for
low star formation efficiencies). In the outer parts of the disk, beyond
the minima of the age profiles, the high efficiency models show inverted
age profiles, while the low efficiency models show declining age profiles.
A more thorough investigation is clearly warranted.

\begin{figure}
\begin{center}
\hspace{0.50cm}
\psfig{figure=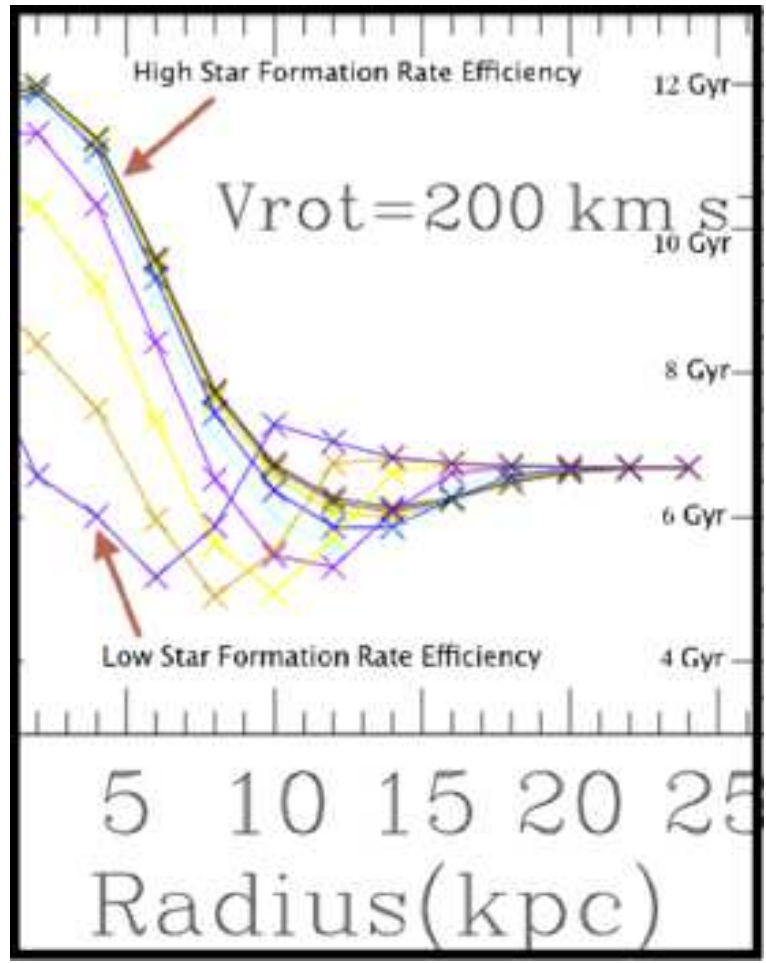,width=10.cm}
\caption{Radial, mass-weighted, age profiles for the fiducial Milky Way
models of Moll\'a \& D\'iaz (2005), for a range of star formation
efficiencies.  All models possess U-shaped age profiles, with the position
and depth of the trough being depending upon the adopted star formation
efficiency.  See text for details.}
\label{fig2}
\end{center}
\end{figure}

\acknowledgments 
Without the help of our collaborators, this work would not have been 
possible; we thank them all for their ongoing advice and guidance. BKG 
thanks both Monash and Saint Mary's Universities for their generous 
visitor support, and the organisers of what was an incredibly exciting,
rewarding, and collegial School and Workshop.  BKG, SC, MM and DC 
acknowledge the support of
the UK's Science \& Technology Facilities Council (ST/F002432/1 \&
ST/H00260X/1). SC acknowledges support 
from the the BINGO Project (ANR-08-BLAN-0316-01).

\end{document}